\documentclass[lettersize,journal]{IEEEtran}
\usepackage{graphicx}
\usepackage{cite}
\usepackage{picinpar}
\usepackage{amsmath}
\usepackage{url}
\usepackage{flushend}
\usepackage[latin1]{inputenc}
\usepackage{colortbl}
\usepackage{soul}
\usepackage{multirow}
\usepackage{pifont}
\usepackage{color}
\usepackage{alltt}
\usepackage[hidelinks]{hyperref}
\usepackage{enumerate}
\usepackage{breakurl}
\usepackage{epstopdf}
\usepackage{pbox}
\usepackage{verbatim}
\usepackage{setspace}

\begin{document}

\title{Haptic Feedback of Tool Vibrations\\Facilitates Telerobotic Construction}

\author{
	Yijie Gong, Haliza Mat Husin, \emph{Member, IEEE}, Ecda Erol, 
	\\ Valerio Ortenzi, and Katherine J. Kuchenbecker, \emph{Fellow, IEEE}
\thanks{This work was partially supported by the Deutsche Forschungsgemeinschaft (DFG, German Research Foundation) under Germany's Excellence Strategy -- EXC 2120/1 -- 390831618.
Yijie Gong, Haliza Mat Husin, and Katherine J. Kuchenbecker are with the Haptic Intelligence Department, Max Planck Institute for Intelligent Systems, 70569 Stuttgart, Germany (email: gong@is.mpg.de, hmh@is.mpg.de, kjk@is.mpg.de). Katherine J. Kuchenbecker is also with the University of Stuttgart.
Ecda Erol and Valerio Ortenzi were previously with the Haptic Intelligence Department, Max Planck Institute for Intelligent Systems, 70569 Stuttgart, Germany. Ecda Erol is with the Department of Mechanical and Process Engineering, ETH Zurich, Zurich, 8092, Switzerland (email: ecda.erol@alumni.ethz.ch).
	}
\thanks{Manuscript received Month xx, 2022; revised Month xx, xxxx; accepted Month xx, xxxx.}
}
\markboth{IEEE}%
{}

\IEEEpubid{This work has been submitted to the IEEE for possible publication. Copyright may be transferred without notice, after which this version may no longer be accessible.}

\maketitle
	
\begin{abstract}
Telerobotics has shown promise in helping workers safely manipulate building components on construction sites; however, its primary reliance on visual feedback limits efficiency in situations with stiff contact or poor visibility. Reliable and economical haptic feedback could fill this perception gap and facilitate construction activities. Thus, we designed an audio-based haptic feedback system that measures the vibrations experienced by the robot tools and enables the operator to feel them in real time. Accurate haptic transmission was achieved by optimizing the positions of the system's off-the-shelf accelerometers and voice-coil actuators. A user study was conducted to evaluate how this naturalistic type of vibration feedback affects the operator's performance in telerobotic assembly. Thirty participants used a bimanual teleoperation system to build a structure under three randomly ordered haptic feedback conditions: no vibrations, one-axis vibrations, and three-axis vibrations. The results show that users took advantage of both kinds of haptic feedback after gaining some experience with the task, causing significantly lower vibrations and forces in the second trial. Subjective responses indicate that haptic feedback reduced the perceived task difficulty, task duration, and fatigue. These results demonstrate that providing this type of vibrotactile feedback on teleoperated construction robots would enhance user performance and experience.
\end{abstract}


\section{Introduction}

\IEEEPARstart{T}{eleoperation} has been shown to increase safety and comfort for human workers in applications such as minimally invasive surgery, search and rescue, and construction~\cite{niemeyer2016}. 
In telerobotic construction activities, the operator usually controls the robot using joysticks and levers, standing at a distance from the end-effector to ensure their own physical safety~\cite{Saidi2016}. In such scenarios, operators rely predominantly on direct vision or camera feeds to operate the robot. Given the chaotic outdoor setting, this visual feedback often suffers from poor viewing conditions and complete occlusion, leading to uncontrolled interactions between the robot and its environment, especially during the contact-heavy process of assembly~\cite{MELENBRINK2020}.
Researchers have thus been working on augmenting the operator's capabilities by equipping robots with sensors~\cite{Cheng2017} and providing operators with haptic feedback~\cite{Culbertson2018haptics}. 

Haptic feedback systems in teleoperation usually measure or estimate the contact force of the robot and display that force to the user in real time~\cite{Kuchenbecker2018}, thereby providing \textit{direct force feedback}. 
Horie et al.~\cite{Horie2001} and Hirabayashi et al.~\cite{HIRABAYASHI2006563} used force sensors to measure the force applied to the robot end-effector and displayed corresponding forces through a PHANToM haptic device and a customized one, respectively. Similarly, Tang et al.\ developed a haptic feedback system for an excavator to provide estimated force feedback to the operator via a motor-driven joystick~\cite{XinxingTang2009}. Their user study showed that this type of force feedback decreased both the applied force and the contact time in a block movement task.
As a drawback, most of these systems are designed to be affixed to a table with a monitor, which is impractical in construction as the operator needs to move around the site. Additionally, three-axis force sensors are typically cost-inefficient for construction. 

Some researchers have also explored \textit{sensory substitution}, i.e., displaying forces through other channels. For instance, Massimino and Sheridan mapped the magnitude of the measured force to that of a fixed-frequency vibration and played that vibration to the user~\cite{MASSIMINO1992109}; their user study showed high success rate and time efficiency in peg-in-hole tasks with this type of vibration feedback. More specifically, in construction scenarios, Nagano et al.\ measured the accelerations of the robot manipulator with a one-axis piezoelectric sensor and played fixed-frequency vibrations through amplitude modulation to the user's wrist using a voice-coil actuator; their user study showed that this type of force feedback decreased the peak force in a bar insertion task~\cite{Nagano2020}.

\IEEEpubidadjcol

\textit{Naturalistic vibrotactile feedback} provides an appealing alternative to direct force feedback and sensory substitution for construction. Generally, vibrations are simpler to measure and output than forces in an unstructured environment, as vibrotactile sensors and actuators are usually small and inexpensive.
Moreover, human vibrotactile perception has a wide bandwidth up to 1000~Hz, which includes the high frequencies that are typically stimulated in contacts between rigid objects being assembled~\cite{BELL199479}. 
This capability has inspired researchers to provide vivid broad-bandwidth haptic feedback during teleoperation. 
In a first effort, Kontarinis and Howe designed a two-fingered teleoperation system with both standard force feedback and naturalistic vibrotactile feedback~\cite{kontarinis1995tactile}; one axis of the vibrations on the robot's fingertip was measured by an accelerometer and played back to the user via an inverted loudspeaker. Their study showed that naturalistic vibrotactile feedback helps reduce the user's reaction time and the force exerted during a puncture task. 

More recently, VerroTouch~\cite{Kuchenbecker2010VerroTouch,McMahan2011} is a naturalistic vibrotactile feedback system that was developed for da Vinci robots (Intuitive Surgical, Inc.), which natively have no haptic feedback. VerroTouch uses three-axis accelerometers to measure the left and right robot tool vibrations and custom voice-coil actuators attached to the da Vinci handles to display the vibrations to the user's hands~\cite{McMahan2011}. The initial version used a single accelerometer axis~\cite{Kuchenbecker2010VerroTouch}, and the later version summed all three axes~\cite{McMahan2011}, but this difference was never investigated. Studies with surgeons and non-surgeons showed strong preferences for having this type of vibrotactile feedback in robotic surgery training tasks~\cite{Koehn2014SurgeonsAN}.   
VerroTouch was created with custom electronics for use on the da Vinci robot, but its approach can be adapted to other use cases, including handheld interfaces with no connection to ground. Compared with vibrations that represent forces through sensory substitution, broad-bandwidth vibrations are more intuitive because they reproduce natural physical contact sensations that humans feel when directly using tools~\cite{kontarinis1995tactile,McMahan2011}.

In this context, the goal of this study is to upgrade VerroTouch and extend its usage for construction scenarios. We aim to design and evaluate an easy-to-make, modular, naturalistic, and reliable vibrotactile feedback system that enables the user to feel the real-time broad-bandwidth vibration that a construction robot's end-effector is experiencing. In addition to the quantitative evaluation of the system performance, we also want to understand the effects this type of haptic feedback has on the operator during telerobotic assembly tasks. We hypothesize that vibrotactile feedback of tool vibrations will  
\begin{enumerate}
\item [\textit{\textbf{H1:}}] help the user create smaller tool vibrations and exert smaller forces on the construction materials, 
\item [\textit{\textbf{H2:}}] increase the user's confidence and decrease their mental stress,
\item [\textit{\textbf{H3:}}] be enhanced by displaying the sum of all three orthogonal tool vibrations rather than a single axis, and
\item [\textit{\textbf{H4:}}] generally lead to a positive correlation between user hand size and preferred feedback magnitude.
\end{enumerate}

To test these hypotheses, we build a haptic system using off-the-shelf audio components for high affordability, accessibility, modularity, and robustness. The system's working mechanism is introduced in Section~\ref{sec:system}.
We systematically select accelerometers, audio equipment, and voice coils for a highly integrated design, and we investigate how to optimally place the accelerometers and the voice-coil actuators to offer high-quality feedback to the user.
Out of safety concerns in real unstructured construction sites, we validate our system's performance with an accurate commercial teleoperation system on a smaller scale. Thus, we use a da Vinci Si robot for this study, as it maps human motion to robot motion naturally and provides no form of haptic feedback. We then conduct a user study to test our hypotheses. Section~\ref{sec:userstudy} describes the methods of this experimental evaluation. The results are presented in Section~\ref{sec:results} and discussed in Section~\ref{sec:discussion}. We conclude with implications, limitations, and future potentials of our work in Section~\ref{sec:conclusion}.

\begin{figure}[t]
    \centering
    \includegraphics[width=\columnwidth]{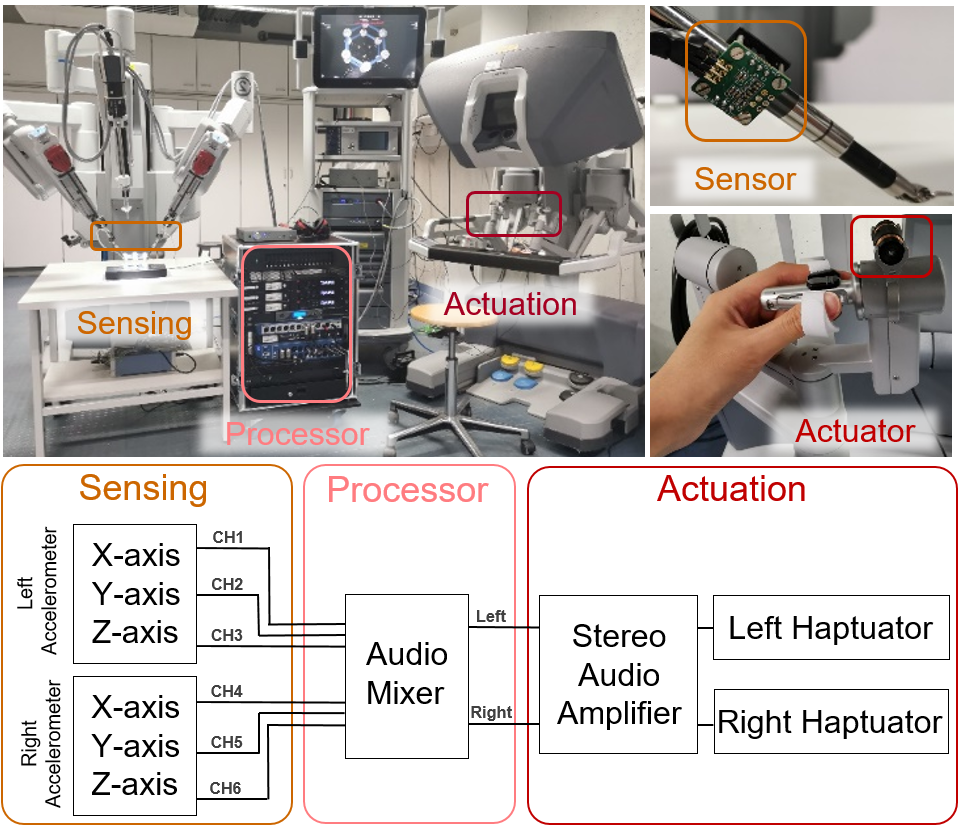}
    \caption{Our audio-based haptic feedback system with a da Vinci robot; it provides real-time vibrotactile feedback to the operator. The system comprises two accelerometers for vibration sensing, off-the-shelf audio equipment for processing (an audio mixer) and amplification (a stereo audio amplifier), and two voice-coil actuators to output the vibrations.}
    \label{fig:system}
\end{figure}

\section{Haptic Feedback System}
\label{sec:system}
Our audio-based vibrotactile feedback system consists of sensing, processing, and actuation units, as shown in Fig.~\ref{fig:system} on a da Vinci Si robot. We were inspired by several other haptic systems that employed audio technology, such as recording VerroTouch vibrations as audio signals~\cite{McMahan13-JRS-Recording}, adapting audio processing techniques to compress vibrotactile signals~\cite{Chaudhari12}, outputting vibrotactile signals with a sound card and an audio amplifier~\cite{Pezent2021}, and generating haptic vibrations with an audio speaker~\cite{Minamizawa2012,Israr2019}.
In detail, our system comprises: two three-axis high-bandwidth accelerometers attached on the two tools of the da Vinci to measure vibrations; an audio mixer that processes the sensed vibrations and outputs them to the actuation unit; and two commercial voice-coil actuators that receive the amplified signals from a stereo audio amplifier and play the sensed vibrations to the user.

\subsection{System Design}
Human vibrotactile perception provides a baseline for the minimal bandwidth requirement: 20--1000\,Hz~\cite{BELL199479}. Thus, we chose components that all have suitable frequency bandwidth to guarantee high-fidelity transmission of the vibration experienced by each tool. 
For the sensing unit, two three-axis accelerometers on evaluation boards (Analog Devices EVAL-ADXL 356b, 20\,mm $\times$ 20\,mm $\times$ 5\,mm, 0--1500\,Hz bandwidth) were selected and attached to the two tools of the robot to capture their vibrations.
For the processor unit, a digital audio mixer (Soundcraft Ui24R) with a band-pass filter set to 80--1000\,Hz mixes, filters, and adjusts the magnitude of the signals from the accelerometers. A slightly narrowed range is used to filter out the low-frequency ego-vibrations of the da Vinci robot and prevent closed-loop instability~\cite{McMahan2011}. 
For the actuation unit, a stereo audio amplifier (Renkforce T21, 20--20\,000\,Hz) with a maximum output power of 50\,dB per channel drives both actuators (Tactile Labs Haptuator Redesign).
The Haptuator (16\,mm diameter $\times$ 29\,mm, $>$1000\,Hz bandwidth) provides high-quality one-axis haptic vibration feedback~\cite{McMahan14-HS-Vibrations}.
All components in our telerobotic system (the da Vinci Si robot and the haptic feedback system) share the same electrical ground to avoid systematic noise.

Placement and orientation of the components have a decisive impact on system performance.
The accelerometer's sensitivity to contacts on the tool end-effector depends on its mounting location, as investigated in Section~\ref{sec:positionAcc}.
Similarly, the Haptuator's ability to accurately vibrate the user's fingers at a broad range of frequencies depends on how it is mounted, so its placement is discussed in detail in Section~\ref{sec:positionActuator}.

\begin{figure}[t]
    \centering
    \includegraphics[width=\columnwidth]{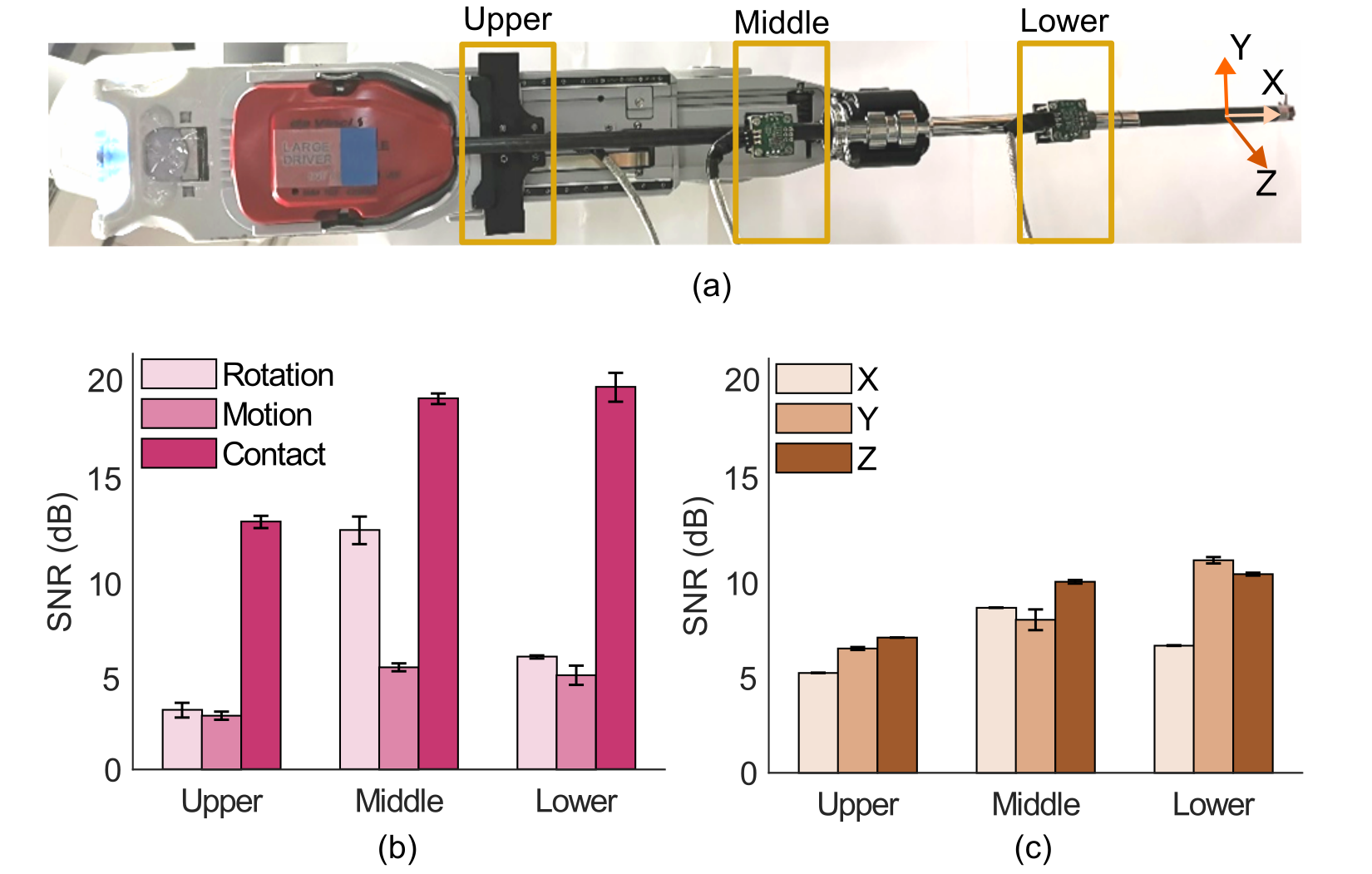}
    \caption{(a) Three possible locations for the accelerometer. (b) The mean SNR of each accelerometer for the three basic actions: rotation, motion, and contact. (c) The mean SNR of each axis of the three accelerometers for contact. Error bars show the standard deviation over trials.}
    \label{fig:snracc}
\end{figure}

\subsection{Positioning the Accelerometer on the Tool}
\label{sec:positionAcc}
Our goal is to provide the user with realistic vibrotactile feedback of the tool contact vibrations and to reduce the transmitted magnitude of internal robot vibrations caused by tool motion and rotation. Regardless of their source, induced vibrations travel along stiff structures with gradually attenuated strengths and are strongly affected by structural resonances.
To determine the optimal accelerometer placement location in our system, custom 3D-printed brackets were used to mount identical accelerometers at three different locations on the left tool of the robot, as shown in Fig.~\ref{fig:snracc}(a). 
The upper location is next to the mounting point of the robot tool, the same location as that of VerroTouch~\cite{McMahan2011}. 
The middle location is on the tool shaft, and the lower location is on the cannula, a steel tube that supports the tool.
The vibration during three teleoperated actions was captured for about 15~s each: (i)~rotation, where the tool rotates around its central x-axis; (ii)~motion, including translation and rotation, where the tool moves in all directions to simulate the reaching process; and (iii)~contact, where the end effector collides with another object. We chose these basic actions as they often occur in assembly tasks, for example when moving and rotating the tool to relocate an object.

\begin{figure}[t]
    \centering
    \includegraphics[width=\columnwidth]{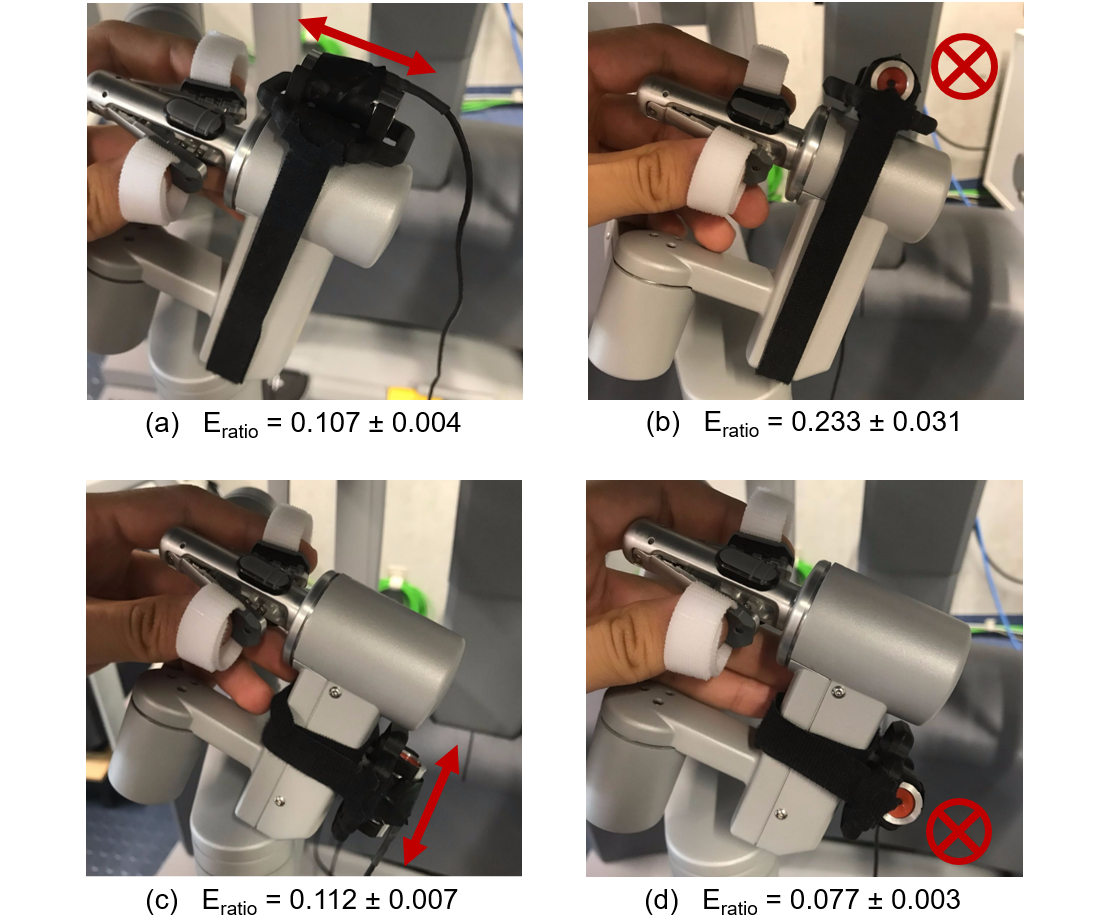}
    \caption{Tested locations of the actuator on the left handle: (a) parallel to the upper part of the handle, (b) perpendicular to the upper part of the handle, (c) parallel to the lower part of the handle, and (d) perpendicular to the lower part of the handle. The arrows and crosses show the direction of the actuator's vibration output in each location. The $E_{\mathrm{ratio}}$ results are shown below each location as mean $\pm$ standard deviation.}
    \label{fig:actu}
\end{figure}

We used signal-to-noise-ratio (SNR, the ratio of the power of a valuable signal to the power of background noise) to evaluate the signal quality in each sensor location. The root mean square (RMS) of the three-axis accelerations was defined as the signal, and the noise was the accelerations from a separate experiment when the tool did not move. 
As shown in Fig.~\ref{fig:snracc}(b), the accelerometers at the middle and lower locations have higher SNR for contact. However, the accelerometer at the middle location measured strong distracting vibrations during rotation and thus has the highest rotation SNR; these vibrations seem to originate from the internal bearing supporting the tool shaft. Therefore, the lower location is the best option for accelerometer placement in this study.

We also analyzed the SNR by axis for the contact action alone, as shown in Fig.~\ref{fig:snracc}(c). The SNRs of the three axes are similar for the upper and middle locations, while the x-axis of the lower location performs worse than the other two axes. These differences likely stem from the structure of the tool.

\subsection{Positioning the Actuator on the Handle}
\label{sec:positionActuator}
The quality of the vibrotactile feedback is limited by the actuator's capacity to replay the tool-contact vibrations on the handle. To provide naturalistic vibrations, distortion and attenuation from the actuators to the hands should be minimal.

We consider four possible locations for the actuator, as shown in Fig.~\ref{fig:actu}.
To compare the quality of the vibration signals, we fed the actuator with the same contact signals and compared the output on a metric based on the acceleration signal energy (ASE) of the vibration source and the vibration generated by the actuator. A three-axis accelerometer was rigidly attached close to where the user's fingertips hold the handle to record the actual vibration feedback. The ASE for one axis is calculated as:
\begin{equation}
    E_{i}=\int_{-\infty}^\infty |a_{i}(t)|^2 dt
    \label{eqn:spectralEnergy}
\end{equation}
where $a_{i}$ is the acceleration on channel ${i}$. The ratio between the sum of the ASE on the handle ($E_{x}$, $E_{y}$, and $E_{z}$) and the acceleration source ($E^{*}_{x}$, $E^{*}_{y}$, and $E^{*}_{z}$) was calculated as:
\begin{equation}
    E_{\mathrm{ratio}}=\frac{E_{x}+E_{y}+E_{z}}{E^{*}_{x}+E^{*}_{y}+E^{*}_{z}}
    \label{eqn:spectralEnergyRatio}
\end{equation}
A higher ratio means better vibrotactile signal transmission to the user's fingers. As shown in Fig.~\ref{fig:actu}, the ratio is clearly highest when the actuator is mounted in position (b). Therefore, we decided to use this location for our study.

\begin{figure}[t]
    \centering
    \includegraphics[width=\columnwidth]{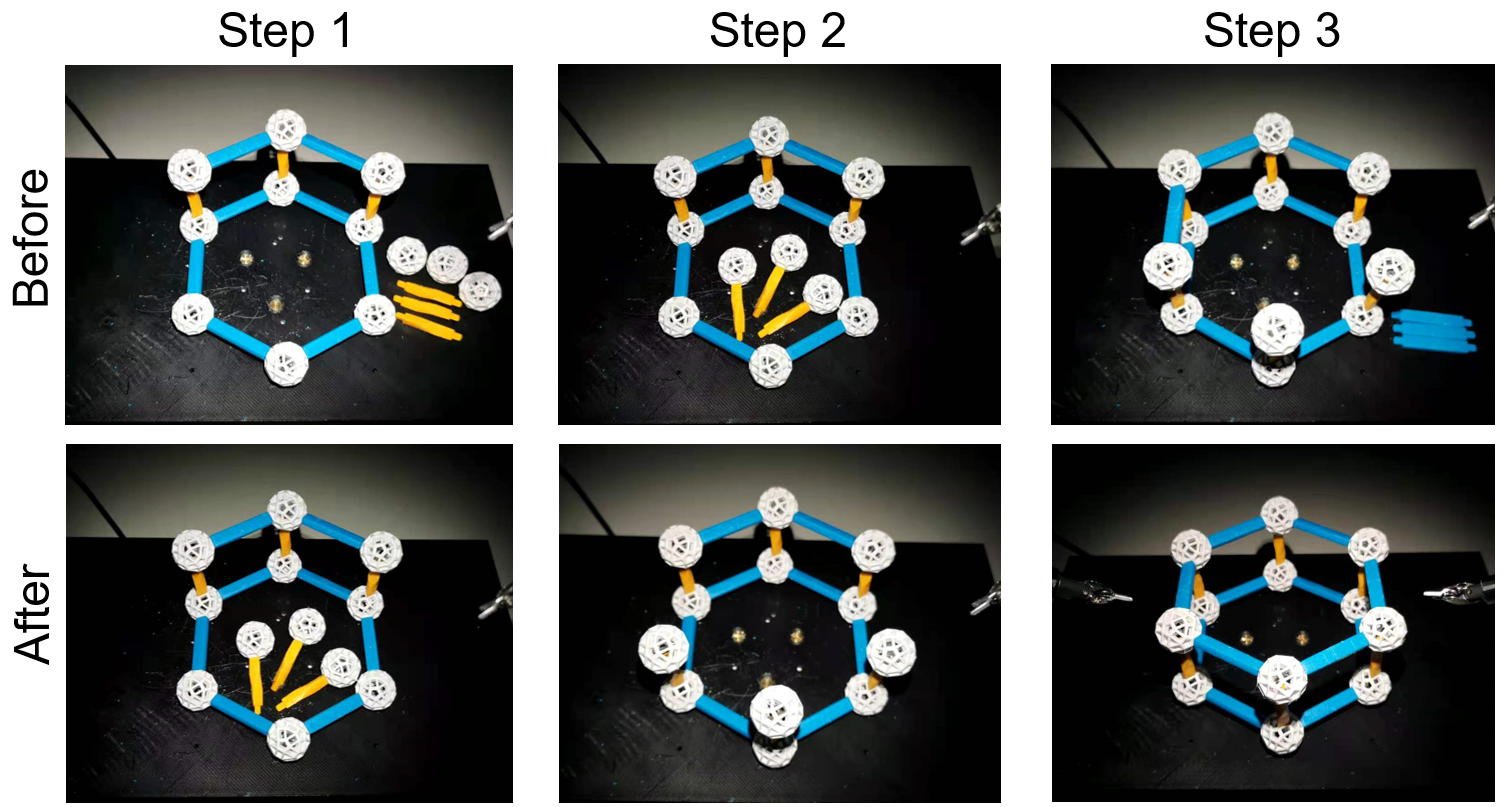}
    \caption{In step 1 of the construction task, the participant assembles three pairs of yellow sticks and white spheres. In step 2, they insert the assembled parts into three base spheres. In step 3, they connect the upper spheres with four blue sticks to form the final structure.}
    \label{fig:steps}
\end{figure}

\subsection{Signal Reproduction from the Tool to the Handle}
We then measured the quality of the reproduced signal to ensure that the user feels the real vibration from the tool. We measured such quality by using four identical accelerometers, which we attached on the two robot tools in the selected lower location and on the handles near the selected location for the actuators. An experimenter teleoperated the robot for the three actions mentioned in Section~\ref{sec:positionAcc}: rotation, motion, and contact. The signals from the accelerometers attached to the tools were processed by the audio mixer, and the accelerometers on the handles measured the vibrations from the actuators. 
Since humans cannot easily perceive the direction of vibrations~\cite{InwookHwang2013}, we convert the three-axis acceleration signals to a one-axis signal by simple arithmetic summing in the time domain, as done in some prior work~\cite{Landin2010,McMahan2011}; this three-to-one function is easily configured in our audio mixer.
As articulated in \textit{\textbf{H3}}, we hypothesize that these summed acceleration signals have more information than any of the individual one-axis signals from the accelerometer, since contact vibrations can occur in all directions; we test this hypothesis in Section~\ref{sec:userstudy}.
The similarity of the summed signals measured on the robot tools and the robot handles during the quality experiment was analyzed using cross-correlation ({\tt xcorr} and {\tt corrcoef} in MATLAB).
The results show that the accelerations on each handle are highly similar to the signal from the corresponding accelerometer ($R_{\textrm{left}} = 0.454$, $R_{\textrm{right}} = 0.463$, $p < 0.001$ and $t_{\textrm{delay}}=0.042$\,s for both), proving that our haptic system reproduces the signal accurately and with minimal time delay.

Compared with VerroTouch, our new audio-based system is easy to build from off-the-shelf components, requires no custom electronics, and allows quick reconfiguration through the mixer. Furthermore, it provides real-time vibrotactile feedback with lower noise and less signal attenuation than VerroTouch due to superior sensing, processing, and actuation. Therefore, we believe this technical approach is capable of providing high-fidelity naturalistic vibration feedback for telerobotic construction and many other applications.

\section{Experimental Evaluation}
\label{sec:userstudy}
We conducted a user study to evaluate how the naturalistic vibrotactile feedback provided by our system affects operator performance during telerobotic assembly tasks.

\subsection{Participants}
Thirty people participated in the study (13 female, 17 male; age $\in [23,40]$, 28.6 (mean) $\pm$ 3.49 (standard deviation)); twenty-six participants had an engineering background. 
When asked on a scale from 0~(novice) to 100~(expert), few were familiar with telerobotics (11.7 $\pm$ 14.6) or haptics research (11.9 $\pm$ 16), while many had some experience with construction activities such as assembling LEGO bricks or IKEA furniture (68.5 $\pm$ 28.3). All gave signed consent and were paid 8 euros per hour for participation. 
All participants were healthy and right-handed, and all reported normal or corrected-to-normal vision. 
The experiment followed procedures approved under the Haptic Intelligence framework agreement from the Max Planck Ethics Council with protocol number F012B.

\subsection{Experimental Setup}
We asked participants to perform a three-step assembly task, as shown in Fig.~\ref{fig:steps}.
Participants need to assemble construction toys (Zometool Inc.) from a half-built state (Fig.~\ref{fig:steps} upper left) to a fully finished state (Fig.~\ref{fig:steps} lower right).
The construction toys comprise sticks and hollow spheres; the assembly procedure includes picking up the parts, moving and inserting parts into other parts, and attaching them to the existing structure, mimicking the on-site construction assembly process. 
The detailed procedure has the following three steps:
\begin{itemize}[\itemindent=10mm]
	\item[Step 1:] connect three pairs of yellow sticks and spheres;
	\item[Step 2:] insert the yellow sticks into three base spheres;
	\item[Step 3:] connect the upper spheres with four blue sticks.  
\end{itemize}
Participants viewed the task materials through the da Vinci Si's stereoscopic endoscope, and they manipulated the task materials with two needle drivers at normal motion scaling. The setup of the camera and tools was standardized, and participants were not allowed to move the camera or clutch the tools.

To validate \textit{\textbf{H1}}, we recorded the vibrations of both tools and used a six-axis force-torque sensor (ATI Mini40) under the construction baseplate to record the forces applied to the task materials.
To evaluate our haptic system's effects on mental stress (\textit{\textbf{H2}}), participants wore a wrist-mounted tracker (Polar OH1) to collect their heart rate during the experiment.
Participants also wore passive noise-canceling headphones to mask ambient sounds.

The study included three different vibrotactile feedback conditions to investigate the impact of our haptic system: without feedback (F0), with the feedback from only one accelerometer axis (F1), and with the feedback computed from summing all three axes of the respective accelerometer (F3).
The F1 and F3 conditions were designed to test our hypothesis \textit{\textbf{H3}} that summed three-axis vibrotactile feedback is more informative than one axis. F1 used the accelerometer's x-axis, which aligns with the tool shaft, as it is less sensitive to the direction of contact than the two other axes.

To validate \textit{\textbf{H4}}, we measured each participant's hand volume using a container filled with water on top of a digital scale~\cite{Hughes2008}. We anticipated that the preferred magnitude would generally be correlated with individual hand mass, as the vibrotactile actuator must be driven with a larger input to achieve the same handle acceleration for a larger hand.

\subsection{Experimental Protocol}
\label{sec:Protocol}
Participants were given five minutes to get familiar with the system settings and tasks (practice section) and were then asked to complete the three-step assembly task within 20 minutes (experimental section).
During the practice section, participants were allowed and encouraged to adjust the system by tuning the vibrotactile feedback magnitude (up to 10\,dB) according to their preference.
Each hand of the participant was measured twice up to the wrist joint, and the mean volume for each hand was recorded.
Then participants repeated the 25-minute-long trial (including practice and experimental sections) under three different haptic feedback conditions (F0, F1, F3).
To control for learning, we randomly assigned the thirty participants to six groups; each group experienced the conditions in one of the six possible presentation orders (F0--F1--F3, F0--F3--F1, F1--F0--F3, F1--F3--F0, F3--F0--F1, F3--F1--F0).

Two questionnaires were administered to collect subjective feedback, as shown in Table~\ref{table_a}. 
The first one was used to record the participant's subjective impression immediately after each trial.
It consists of four customized questions (Q1--Q3, Q5) and six questions about workload, directly adapted from the NASA task load index (TLX, Q4.1--Q4.6): they focus on mental demand, physical demand, temporal demand, performance, effort, and frustration~\cite{HART1988139}.
Each question has a scale from 0 (not at all) to 100 (completely).
The second questionnaire was designed to compare performance across trials and was administered at the end of the experiment.
It includes three free-response questions (Q6, Q11, Q12) and four questions regarding preference among the three trials (Q7--Q10). 

\begin{table}[!t]
	\renewcommand{\arraystretch}{1.3}
	\caption{Questionnaires completed during the study}
	\centering
	\label{table_a}
	\renewcommand{\arraystretch}{1.1}
	\resizebox{\columnwidth}{!}{
		\begin{tabular}{c l}
			\hline\hline \\[-3mm]
			& \multicolumn{1}{c}{Post-trial questions}\\[.2ex] \hline

			Q1 &   How confident were you in moving the building materials? (0--100)\\
			
			Q2 &   How confident were you in assembling the structure? (0--100)\\
			Q3 &   How realistic was your interaction with the building materials? (0--100)\\
			Q4.1 &   How mentally demanding was the task? (0--100)\\
			Q4.2 &   How physically demanding was the task? (0--100)\\
			Q4.3 &   How hurried or rushed was the pace of the task? (0--100)\\
			Q4.4 &   How successful were you in accomplishing what you were asked \\\smallskip
			& to do? (0--100)\\
			Q4.5 &   How hard did you have to work to accomplish your level of \\\smallskip
			& performance? (0--100)\\
			Q4.6 &   How insecure, discouraged, irritated, stressed, and annoyed were \\\smallskip
			& you? (0--100)\\
			Q5 &  What comments do you have about your experience performing \\\smallskip
			& this construction session? (text)\\[1ex]

			\hline \\[-3mm]
			 & \multicolumn{1}{c}{Post-experiment questions} \\[.2ex] \hline
			Q6 & 	What differences did you notice, if any, between three construction \\\smallskip
			& sessions? (text)\\
			Q7 &   Which construction trial was the fastest to complete? (1, 2, or 3)\\
			Q8 &   Which construction trial was the easiest to complete? (1, 2, or 3)\\
			Q9 &   Which construction trial was the most fatiguing? (1, 2, or 3)\\
			Q10 &   Which construction trial did you like best? Why? (1, 2, or 3; text)\\
			Q11 &   For the two sessions that included this step, how did you decide \\\smallskip
			 & what gain level to choose? (text)\\
			Q12 &   Do you have any additional comments about the study? (text)\\[1ex] 
			\hline\hline
		\end{tabular}
	}
\end{table}

\begin{figure}[t]
    \centering
    \includegraphics[width=\columnwidth]{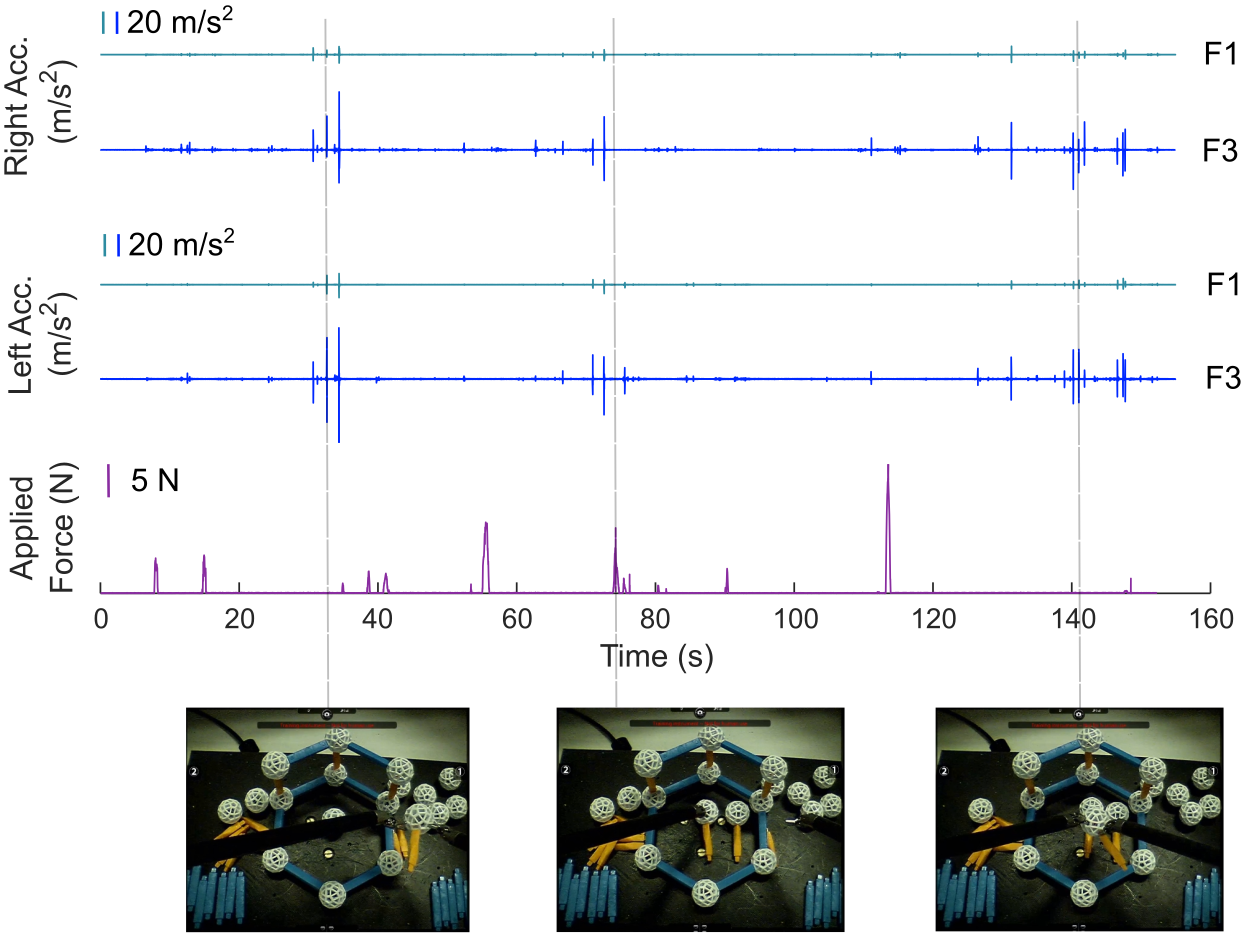}
    \caption{Tool vibration and applied force magnitude for step 1 of one trial in the study. Each tool acceleration shows the signals of both F1 and F3. The screenshots at the bottom show the participant's left eye view of the task materials at the three indicated time points.}
    \label{fig:signals}
\end{figure}

\subsection{Sample Trial Data}
Fig.~\ref{fig:signals} shows sample acceleration and force data from one participant performing step 1 of the assembly task. We depict both the F1 and F3 signals for both tools and include three screenshots of the participant's left eye view for context. The supplementary video associated with this article shows this same interaction along with animated plots of the corresponding left tool vibration, right tool vibration, and applied force magnitude over time. 

\begin{figure*}[tp]
    \centering
    \includegraphics[width=\textwidth]{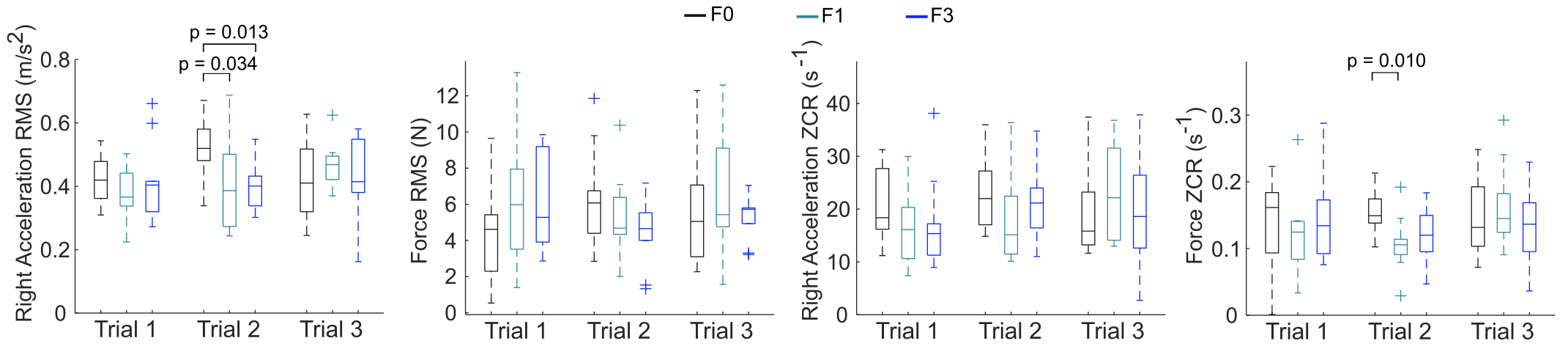}

    \caption{The accelerometer and force data analyzed by trial. (a) RMS of summed acceleration from right accelerometer (dominant hand); (b) RMS of force; (c) ZCR of right acceleration; (d) ZCR of force. The bottom and top of the box represent the 25\% and 75\% responses, and the center line is the median. The + marks indicate outliers. The lines extending past the boxes show the farthest data points not considered outliers. Statistically significant pairwise differences are marked with corresponding p-values. }
    \label{fig:sensordata}
\end{figure*}

\section{Results}
\label{sec:results}
We used the accelerometers on the cannulas to measure the magnitude of the tool vibrations,
the force sensor to measure the applied force,
the task completion time to evaluate the assembly efficiency,
the heart rate (HR) tracker to evaluate the operator's mental stress,
the chosen haptic feedback gains to check the effects of the two versions of haptic feedback,
and the questionnaire responses to evaluate the overall system performance.

\subsection{Signal Processing and Evaluation Metrics}
We set thresholds for acceleration (0.3\,m/s$^{2}$) and force vector magnitude (0.2\,N) to remove baseline noise.
The data above the threshold were retained, and the data below were set to 0.
The RMS and zero-crossing-rate (ZCR, the number of times per second the signal crosses zero) of the thresholded acceleration and force signals were used to quantify the signal's strength and the average interaction intensity.
We evaluated the completion level of each trial but did not analyze this variable in detail because 85\% of the structures were fully completed, and the incomplete structures all happened in the first or second trial.  

We first used Spearman's correlation coefficient to analyze the effects of interactions between the performance metrics. 
We then performed the following quantitative analyses to evaluate the effect of haptic feedback:
\begin{itemize}
    \item Friedman test to analyze the disparity between three feedback conditions; if it is significant, we use Wilcoxon signed-rank tests for post-hoc analysis on the three pairs of two conditions;
    \item Kruskal-Wallis test to analyze the effect of three feedback conditions within each trial; Mann-Whitney test for the post-hoc analysis on three pairs of two conditions within each trial.
\end{itemize}
We used $\alpha = 0.05$ to determine statistical significance, and the Bonferroni-Holm correction was applied to correct the significance levels for multiple comparisons.

\subsection{Correlations Between Metrics} 
\label{subsec:Correlations Between Metrics}
Spearman's correlation was performed on combined data from the three trials ($N = 90$ trials) to evaluate how the tool acceleration metrics relate to the force metrics, mean HR, completion time, hand volume, and TLX workload. When two metrics are positively correlated, one metric increases as the other increases; a negative correlation means one metric increases while the other decreases.  

We found all the metrics for acceleration and force are highly positively correlated ($p < 0.015$ for all), except for force RMS and acceleration ZCR of both tools. 
The acceleration RMSs of both tools have significant positive correlations with the mean HR ($p < 0.001$ for both) and significant negative correlations with completion time ($p < 0.01$ for both).
The acceleration ZCRs of both tools have positive correlations with mean HR ($p < 0.001$ for both), and they have significant negative correlations with completion time ($p < 0.001$ for both).
User hand volume is positively correlated with chosen feedback gain $(p = 0.002)$ and negatively correlated with completion time ($p = 0.006$). 
The TLX workload is positively correlated with completion time ($p = 0.012$) and negatively correlated with acceleration ZCR of the left tool ($p = 0.014$). 
No significant correlations were found in the rest of the metric pairs.

\subsection{The Effect of Haptic Feedback} 
\label{subsec:The effect of haptic feedback}
\subsubsection*{General effect of the three feedback conditions} We combined the data across the three trials to evaluate the effects of the three feedback conditions on the left and right tool accelerations, force, completion time, mean HR, and post-trial questionnaires. There is a significant main effect of feedback condition on the right tool acceleration RMS ($p = 0.014$). Post-hoc analysis shows F0 has a significantly higher RMS acceleration than F1 ($p = 0.036$), and F0 also tends to be higher than F3 ($p = 0.052$). There is no significant main effect of feedback condition in any other signal-based metrics. For the post-trial questionnaires, there is a significant main effect of feedback condition in realistic interaction ($p < 0.001$), and F1 and F3 are significantly more realistic than F0 in post-hoc analysis (F0 vs. F1: $p < 0.001$; F0 vs. F3: $p = 0.004$). No significant differences were found in any other questions.

\subsubsection*{Effect of the three feedback conditions within each trial} Since the task was repeated three times, participants gained experience over time. To reduce the influence of this learning process, we also analyzed the metrics for each trial separately. Fig.~\ref{fig:sensordata} shows the effect of feedback condition on the right tool accelerations and force variables by trial. In trial 1, there is a significant main effect of haptic feedback only on completion time ($p = 0.047$, not plotted). Post-hoc analysis shows that the completion time of F0 has a trend to be higher than that of F3 ($p = 0.051$). No significant differences were found in other metrics in trial 1. 
In trial 2, there is a significant main effect of feedback condition on the right tool RMS acceleration ($p = 0.027$) and force ZCR ($p = 0.003$). Post-hoc analysis shows the right tool acceleration RMS and force ZCR of F0 are significantly higher than those of F3 ($p < 0.05$ for both), and the right tool acceleration RMS of F0 is also significantly higher than that of F1 ($p = 0.013$). No other significant differences were found in trial 2. 
In trial 3, F1 and F3 both trend to give users significantly more confidence in moving the parts ($p = 0.053$) and significantly more realistic interaction with the building materials ($p = 0.064$) compared to F0. No significant differences were found in other metrics for trial 3.

\begin{figure}[tp]
    \centering
    \includegraphics[width=\columnwidth]{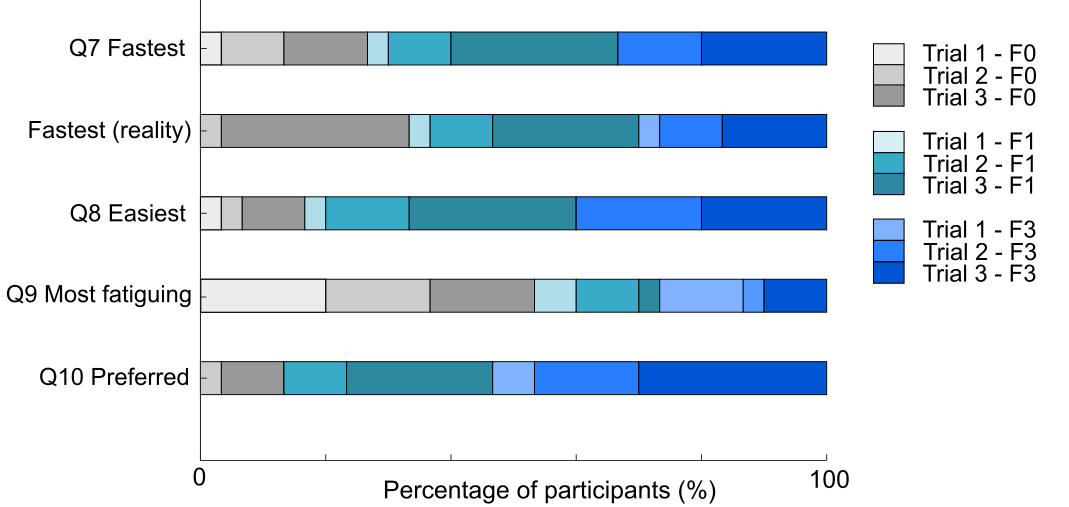}
    \caption{The results of Q7 to Q10. The participant was asked to identify the fastest trial, the easiest trial, the most fatiguing trial and the trial they like the most. The second graph shows the real fastest trial calculated from the recorded completion time.
}
    \label{fig:ques-result}
\end{figure}

\subsubsection*{Post-experiment questionnaire} 84\% of the participants said they could tell the difference between with and without haptic feedback in Q6, while only 6\% of them mentioned that the two haptic feedback conditions felt different. We compared the participants' responses for Q7 to Q10 with their assigned feedback conditions. The results of these four questions are shown in Fig.~\ref{fig:ques-result}. 77\% of the participants chose a trial with haptic feedback (F1 or F3) as the fastest to complete. Surprisingly, based on the measured completion time, almost half (47\%) of the participants did not choose their actual fastest trial as the fastest completed trial. Trial 3 with no haptic feedback had the highest number of unmatched answers. For easiest trial, 84\% of the participants chose a trial with haptic feedback. For most fatiguing trial, only 54\% chose a trial with haptic feedback. For favorite trial, 87\% of the participants chose one with haptic feedback. Their reasons are diverse, such as the feedback helps provide a more realistic experience; it is helpful for beginners to understand the assembly strategies; and it helps when visual feedback is insufficient. Several participants also commented that haptics is usually taken for granted and undervalued during their daily lives.

For the open-ended Q11 about gain selection, 50\% of the participants said they chose the gain to be neither too strong nor too weak, and 17\% of the participants said they set the gain to be at a comfortable level. 23\% of the participants said they changed their preference during the experiment, as they preferred strong vibration initially, and then they realized that somewhat weaker vibration is better and turned it down in the subsequent trial. In the free comments, 10\% of the participants said they did not find the vibrotactile feedback to be helpful, because they found it distracting or strongly relied on visual feedback. Several participants mentioned that they did not realize the importance of haptic feedback until they experienced conditions with and without it.

\subsubsection*{Gain selection} We also compared the difference between the gains chosen by the participants for the two haptic feedback conditions using a paired t-test. The gain for F1 is significantly smaller than F3 ($p = 0.010$). The mean difference is 3.8 dB (F1: 3.64 $\pm$ 6.98; F3: -0.22 $\pm$ 5.94), which means the chosen amplification of one-axis feedback was 2.43 times higher of that of three-axis feedback. We believe that most participants chose gains that gave the two trials similar vibration magnitudes for the same physical contact, which the comments from Q11 also support. 
For F1, the chosen gain shows a significant positive correlation with the measured hand volume ($p = 0.006$). The gain chosen for F3 also positively correlates with hand volume, but not significantly ($p = 0.175$). 

\section{Discussion}
\label{sec:discussion}
Overall, the results from this study strongly support our hypotheses that naturalistic haptic feedback of tool vibrations has positive effects on the performance of assembly activities in both objective (\textit{\textbf{H1}}) and subjective (\textit{\textbf{H2}}) evaluations. However, the highly similar results found for the one-axis and summed three-axis feedback do not support \textit{\textbf{H3}}. Finally, the positive correlation between the magnitude of the haptic feedback and the gain choice supports \textit{\textbf{H4}}.

\subsection{Objective Evaluation} 
During construction assembly activities, strong contact vibrations and forces would deform or damage the structure being built; buildings with smaller construction-induced deformations last longer~\cite{atkinson2003construction}. Thus, we regard it as better performance when participants assemble the components with smaller generated tool vibrations and baseplate forces. 

The objective analysis of our study supports \textit{\textbf{H1}}.
Providing naturalistic haptic feedback of tool vibrations generally reduced the vibrations and forces that the user caused while assembling the structure. We observed that after participants gained experience with the setup through the first trial, they caused lower vibrations and lower forces when they had haptic feedback: specifically, we found lower values for four calculated metrics (RMS and ZCR for right-tool accelerations and forces) in F1 and F3 in trial 2 (Fig.~\ref{fig:sensordata}). 
Moreover, the positive correlation between tool accelerations and contact forces presented in Section~\ref{subsec:Correlations Between Metrics} reinforces this connection. This result is consistent with literature in that both tool acceleration RMS and exerted force are lower when naturalistic vibrotactile feedback is provided~\cite{McMahan2011,McMahan11-HSMR-Force}.
Thus, we can conclude that haptic feedback helps decrease these adverse effects during assembly. The lack of objective differences between F1 and F3 give no support for \textit{\textbf{H3}}.

It is interesting that we do not see a significant effect of haptic feedback in those four metrics (RMS and ZCR for accelerations and forces) in trial 1 or trial 3. 
In trial 1, regardless of the feedback condition, users tended to be more cautious as it was their first time being exposed to the robot and task. 
The participants who did not have haptic feedback commented that they had no concept of the robotic tool strength, and they sometimes damaged the toy structure.
In contrast, the participants who received haptic feedback at the start said they had a more realistic experience and were motivated to complete the task. 
Based on their comments, they could tell the robot's strength from the magnitude of the vibrations during the task. In trial 3, the metric values were similar in all feedback conditions. 
The corresponding comments indicated that participants relied on the experience they had gained from previous trials, hinting that the benefits of past exposure to naturalistic vibrotactile feedback may persist when the feedback is removed. Longer studies are needed to investigate this idea.

\subsection{Subjective Evaluation} 
The mental health of workers is crucial in construction. Due to the dangerous working environment and the high workload, workers often feel stress and anxiety~\cite{Abdelhamid2002}, which may increase accident rates~\cite{langdon2018construction,leung2010impacts}. 
We envision that naturalistic feedback of the vibrations experienced by teleoperated construction robots could help create a lower-pressure and therefore safer environment for workers. 

The questionnaire responses support \textit{\textbf{H2}} about the subjective benefits of this type of haptic feedback: most participants thought they had more realistic interactions with the building materials and that they finished the task faster, more easily, and with less fatigue when they had haptic feedback. 
Furthermore, most of the participants preferred having feedback available when doing assembly activities; this finding is similar to a previous study of VerroTouch, where both surgeons and non-surgeons preferred haptic feedback in surgical training tasks~\cite{Koehn2014SurgeonsAN}. 
According to Section~\ref{subsec:The effect of haptic feedback}, the result of the fastest trial (Q7) is mildly surprising: a group of the participants thought they finished the task most quickly when they had haptic feedback, rather than realizing that their final trial without haptic feedback was actually fastest. 
Thus, we believe that this form of haptic feedback may accelerate time perception. 
Another supportive result is that more than two-thirds of participants thought the trials with haptic feedback were easier~(Q8). 
Our results about fatigue (Q9) also show the positive mental effects of haptic feedback. Furthermore, as described in Section~\ref{subsec:Correlations Between Metrics}, the calming effect of haptic feedback is also supported by the positive correlation between mean~HR and both acceleration RMS and ZCR, showing haptic feedback tends to reduce tool vibrations, and users had lower heart rate on trials with lower tool vibrations. 

Interestingly, we did not observe any differences between F1 and F3 from the questionnaires, indicating that our hypothesis \textit{\textbf{H3}} is not supported. 
The majority of the subjects could discern only whether there was haptic feedback, and they could not differentiate between the two types. 
We believe that their focus on the task may have prevented them from noticing subtle properties of the feedback. 
Moreover, for the long, thin robot tools of the da Vinci, the magnitude difference between the three axes is relatively small, as the contact vibration on one axis propagates easily to the other two axes, and contact vibrations often occur along more than one axis. 
Thus, we can conclude that one-axis feedback seems just as good as three-axis feedback for this robot and task. More investigation is needed to determine whether a difference would ever be perceived by users.
Furthermore, we observed that hand volume was positively correlated to the gain choice, supporting our hypothesis (\textit{\textbf{H4}}) that individuals with larger hands will generally need stronger vibrotactile output; however, large variations were seen across individuals, most likely due differing tactile sensitivity and personal preferences.

\section{Conclusion}
\label{sec:conclusion}

This paper presented a reliable haptic feedback system that uses off-the-shelf audio equipment to provide users with naturalistic real-time vibrotactile feedback while teleoperating a robot. 
The system consists of two accelerometers, an audio mixer, a stereo audio amplifier, and two voice-coil actuators. 
After showing how to optimize sensor and actuator placement, we conducted a user study to evaluate this system and explore the effects of haptic feedback during telerobotic assembly. 
Thirty participants used a da Vinci robot to assemble a structure in three randomly ordered haptic feedback conditions: no feedback, one-axis feedback, and summed three-axis feedback. 
We saw a significant effect of haptic feedback when participants had some experience with the robot and the task: namely, participants exerted less force and produced smaller tool vibrations in trial 2, supporting \textit{\textbf{H1}}. 
Moreover, most participants showed a strong preference for haptic feedback in their qualitative evaluations, believing that it makes the task easier and less fatiguing, supporting \textit{\textbf{H2}}. Interestingly, participants could not distinguish between one-axis and three-axis feedback, undermining \textit{\textbf{H3}}. Participants with larger hands generally chose higher gains for the haptic feedback, supporting \textit{\textbf{H4}}.
This paper's findings about the objective and subjective benefits of this form of haptic feedback provide good guidance to support its future use in teleoperated construction robots, including optimal sensor and actuator placement strategies.
Thus, future work will apply our findings on real construction robots with more realistic tasks in a large workspace. Since it is not convenient to have long cables between the end-effector of a construction robot and the human operator, we plan to further upgrade our system with wireless communication for both sensing and actuation.

\section*{Acknowledgments}
The authors thank the International Max Planck Research School for Intelligent Systems (IMPRS-IS) for supporting \mbox{Yijie} Gong. The authors also thank Bernard Javot and Joey Burns for technical support and Mayumi Mohan, Ravali Gourishetti, and Huanbo Sun for suggestions on this article.



\begin{IEEEbiography}[{\includegraphics[width=1in,height=1.25in,clip,keepaspectratio]{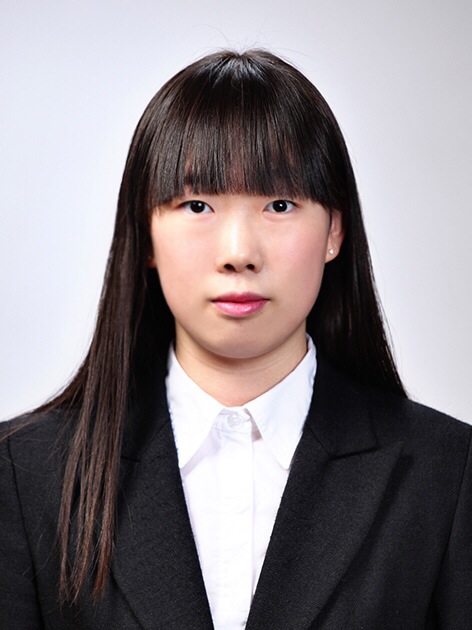}}]
{Yijie Gong} earned a B.Eng.\ degree in Mechatronics Engineering at Northwestern Polytechnical University, China, in 2016 and a M.Eng.\ degree in Mechanical Engineering at Beihang University, China, in 2019. Now she is a doctoral student at the Max Planck Institute for Intelligent Systems in the Haptic Intelligence Department in Stuttgart, Germany, and she is part of the Cluster of Excellence on Integrative Computational Design and Construction for Architecture (IntCDC). Her research mainly focuses on haptics and human-robot interaction.
\end{IEEEbiography}

\begin{IEEEbiography}[{\includegraphics[width=1in,height=1.25in,clip,keepaspectratio]{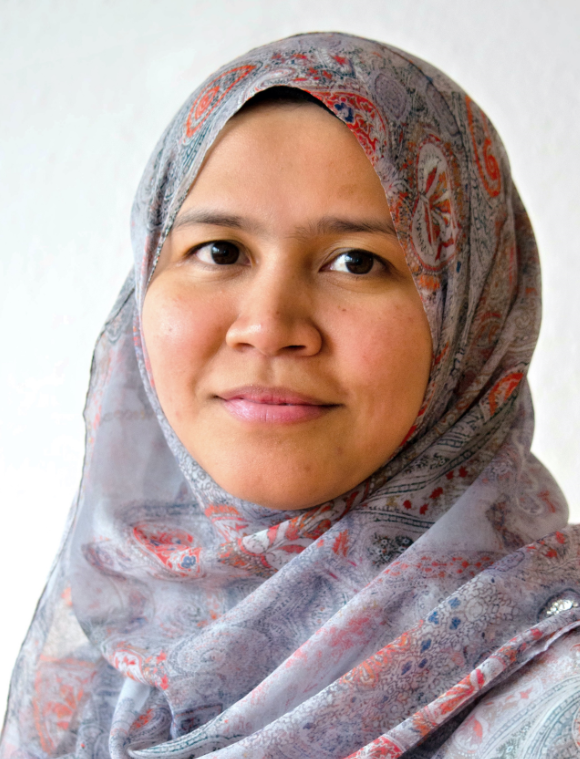}}]
{Haliza Mat Husin} obtained the B.Eng.\ degree in electrical engineering from the University of Malaya, Malaysia in 2002, M.Eng.\ in electrical and telecommunications from the University of Technology Malaysia, Malaysia, in 2005, and M.Sc., in biomedical engineering from the University of Luebeck, Germany, in 2012. She earned her doctorate in Neuroscience in 2020 from the University of Tuebingen, Germany. She is currently working as a Research Scientist in the Haptic Intelligence Department at the Max Planck Institute for Intelligent Systems. Her research interests include haptic interfaces for teleoperation, human-robot interaction, human perception, and physiological signals related to haptic feedback.
\end{IEEEbiography}

\begin{IEEEbiography}[{\includegraphics[width=1in,height=1.25in,clip,keepaspectratio]{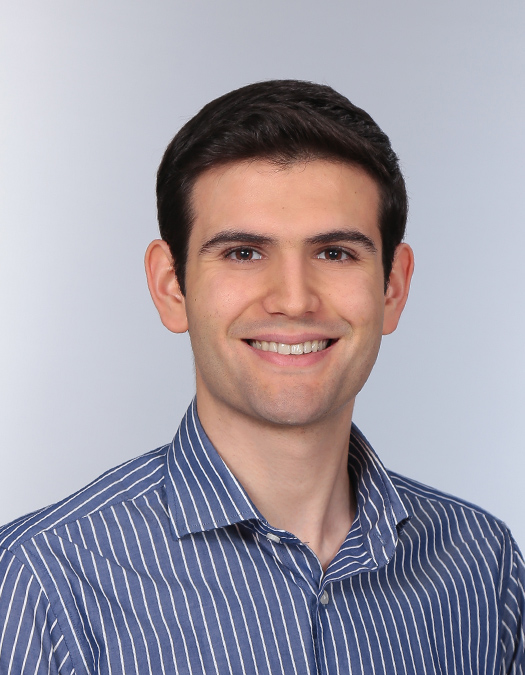}}]
{Ecda Erol} received his B.Sc.\ degree in Mechanical Engineering from Bogazici University, Turkey, in 2019 and his M.Sc.\ degree in Mechanical Engineering from ETH Zurich, Switzerland, in 2022. He worked as a summer intern in the Haptic Intelligence Department at the Max Planck Institute for Intelligent Systems in 2018. After his formal education, he started developing six-axis force-torque sensors for robotic manipulation as a Mechanical Design Engineer at Bota Systems AG in Switzerland. His research focuses on design and modeling of mechanical systems for robotic manipulation.
\end{IEEEbiography}

\begin{IEEEbiography}[{\includegraphics[width=1in,height=1.25in,clip,keepaspectratio]{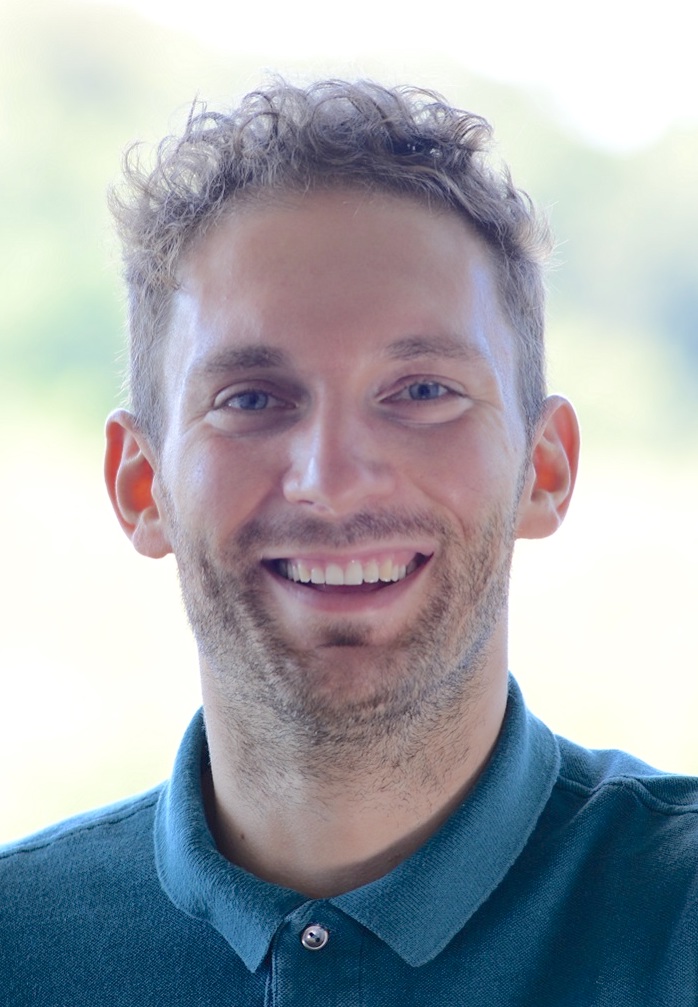}}]
{Valerio Ortenzi} qualified with a B.Sc.\ in Automation and Automated Systems Engineering from Sapienza University of Rome, Italy, in 2010. He went on to study for a M.Sc.\ in Artificial Intelligence and Robotics Engineering, 2012, from Sapienza University of Rome, Italy, and then a Ph.D.\ in Robotics in 2017, from the University of Birmingham, UK. He joined the Centre of Excellence in Robotic Vision at Queensland University of Technology, Brisbane, Australia, as a Research Fellow with Dist.\ Prof.\ Peter Corke. From October 2018 to April 2020, he was a postdoctoral researcher at the University of Birmingham, UK. He worked as a Research Scientist in the Haptic Intelligence Department at the Max Planck Institute for Intelligence Systems from 2020 to 2021. His research interests focus on human-robot interaction and robotic manipulation. New directions of his work go into neuroscience, the psychology of human grasping, and robotic vision.
\end{IEEEbiography}

\begin{IEEEbiography}[{\includegraphics[width=1in,height=1.25in,clip,keepaspectratio]{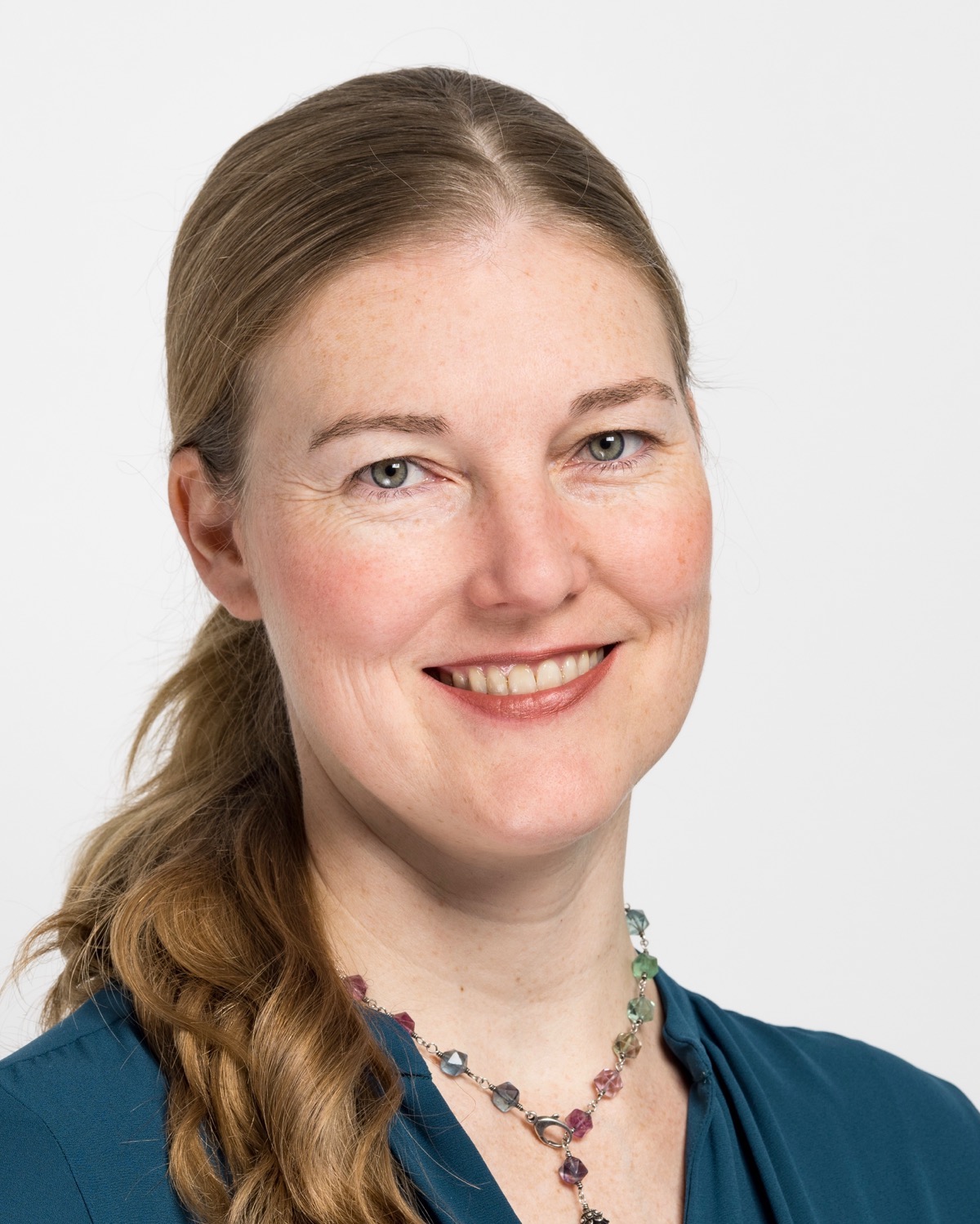}}]
{Katherine J. Kuchenbecker} received the B.S., M.S., and Ph.D.\ degrees from Stanford University, Stanford, USA, in 2000, 2002, and 2006, all in mechanical engineering. After postdoctoral research at the Johns Hopkins University, she was a faculty member in the School of Engineering and Applied Sciences at the University of Pennsylvania, Philadelphia, USA, for 9.5 years before moving to the Max Planck Society in 2017. She currently directs the Haptic Intelligence Department at the Max Planck Institute for Intelligent Systems, Stuttgart, Germany. She is a principal investigator within the Cluster of Excellence on Integrative Computational Design and Construction for Architecture (IntCDC) and an Honorary Professor at the University of Stuttgart. Her research interests include haptic interfaces, haptic sensing systems, and human-robot interaction. She has been honored with a 2009 NSF CAREER Award, the 2012 IEEE RAS Academic Early Career Award, a 2014 Penn Lindback Award for Distinguished Teaching, elevation to IEEE Fellow in 2022, and various best paper, poster, demonstration, and reviewer awards. She co-chaired the IEEE RAS Technical Committee on Haptics from 2015 to 2017 and the IEEE Haptics Symposium in 2016 and 2018.
\end{IEEEbiography}
\end{document}